# Measuring Structural Restraint in an Amorphous Material Using the Variance of Speckle Intensity

Charlotte F. Petersen[1] and Peter Harrowell[2]

[1] *School of Chemistry, University of Melbourne, Melbourne, Victoria 3010, Australia*

[2] *School of Chemistry, University of Sydney, Sydney, New South Wales 2006, Australia*

[1] charlotte.petersen@unimelb.edu.au

[2] peter.harrowell@sydney.edu.au

Abstract

An experimentally accessible measure of structural restraint in supercooled liquids is defined based on the variance of the intensity of speckle pattern arising from elastic scattering of coherent radiation. Using simulation data, the resulting measure of structural restraint is shown to correlate with the temperature dependence of the shear modulus and the Debye-Waller factor. The paper includes consideration of the influence of the choice of the scattering wavevector and the degree of coarse graining used in intensity measurements on the measured restraint.

## 1. Introduction

A liquid, as it is supercooled, exhibits only a small continuous change in the amplitude of the scattered X-rays while undergoing a dramatic mechanical transformation from a fluid to an amorphous solid [1]. This is the essential puzzle posed by the glass transition – how to find some sort of conceptual accommodation for a solidification that cannot be directly attributed to a change in observed structure. There is no shortage of efforts to make this accommodation. A substantial body of literature is based on the proposition that the temperature dependence of the liquid structure is dominated by the accumulation of a small set of stable ('favored') local coordination geometries [2-4]. These favored local structures are invoked to account for both the kinetic stability of the supercooled liquid with respect to crystallization [3] and the mechanical stability of the glass with respect to an applied stress [2]. There are a number of difficulties with this line of thinking: amorphous materials are typically characterised by a large number of contributing local structures, and there is a real problem in establishing any causal connection between a structure and the properties it is supposed to account for. Recently [5,6], we proposed that the problem of connecting amorphous structure to solidity can be directly addressed by expanding our description of structure to include measures of atomic restraint. The essential idea is that it is not the structure itself but its capacity to restrain the motion of the constituent atoms that undergoes a dramatic change on cooling.

An appealing feature of structural restraint as an order parameter is the real possibility of measuring a related quantity experimentally. The method used to define restraint in simulations [5,6] is not experimentally accessible. In ref. [5] we defined structural restraint based on the equilibrium mean squared displacements for each particle calculated using a harmonic approximation of the potential at an instantaneous configuration. This approach provides a measure of restraint that does not require dynamical averaging, a requirement to conceptually justify the description 'structural'. In this paper we shall explore a method for measuring structural restraint using the temperature dependence of the statistics of intensity fluctuations using coherent scattering. The structure factor at a wavevector $\vec{q}$ , collected over a measurement time τ, can be written as

$$S(\vec{q}) = 1 + \frac{1}{N}\sum_{j \neq k} \left\langle \exp\left[-i\vec{q}\cdot(\vec{r}_j - \vec{r}_k)\right]\right\rangle_\tau \qquad (1)$$

where $\langle ... \rangle_\tau$ corresponds to an average over the time interval τ. $S(\vec{q})$ is proportional to the scattered intensity $I(\vec{q})$ measured over the same time interval. The scattering from a disordered arrangement of atoms can give rise to a granular distribution of intensities known as 'speckle'[7]. A speckle pattern contains information about short and intermediate range order as well as a random contribution arising from uncorrelated structure. Both contributions are highly sensitive to particle motion. We can project the scattering onto a plane corresponding to the x and y components of the scattering vector $\vec{q}$ as shown in Fig. 1. With increasing temperature, the amplitude of particle motion increases and the measured intensity of reflection converges to the mean value for a given magnitude q of the scattering vector and the angular variance of the intensity vanishes, resulting in uniform Debye-Scherrer rings [8].

The size of the 'grains' of a speckle pattern are proportional to λz/L where λ is the scattering wavelength, z is the distance between the sample and the detector and L is the radius of the scattering area [9]. Increasing the dimension L of the scattering volume reduces the variation in scattering as measured in a fixed pixel size in the reciprocal space. This observation is the basis of what is now called variable-resolution fluctuation electron microscopy [10], a technique first described by Treacy and Gibson [11]. Working at a fixed scattering wavevector $\vec{q}$, the dependence on Q , the radius of the aperture in diffraction space, of the variance of the scattering intensity

$$V(\vec{q},Q) = \frac{\langle S^2(\vec{q},Q)\rangle}{\langle S(\vec{q},Q)\rangle^2} - 1 \ , \qquad (2)$$

averaged over multiple scattering volumes, provides a measure of the length scale of medium range order in an amorphous sample.

While, in simulations, it is possible to measure properties of a single configuration, experimental measurements of structure take place over a time interval so that measured structures are functions of the measurement time τ. In modelling the scattering from an amorphous sample, we include this averaging over a time interval τ, albeit one much shorter than that used in actual experiments (see Section 2 for details). If the measurement time exceeds the characteristic time for structural relaxation, the averaged scattering intensity will not show a speckle pattern. This characteristic time can be determined from the decorrelation of the intensity. Experimental realisations of such measurements have been carried out using lasers on colloidal glass formers [12] (i.e. dynamic light scattering) and, on atomic glasses, using coherent X-rays in a technique called X-ray photon correlation spectroscopy [13]. The spatial variation of the intensity decorrelation, measured using electron scattering, can be measured using electron correlation microscopy, a method developed by He et al. [14] in 2015.

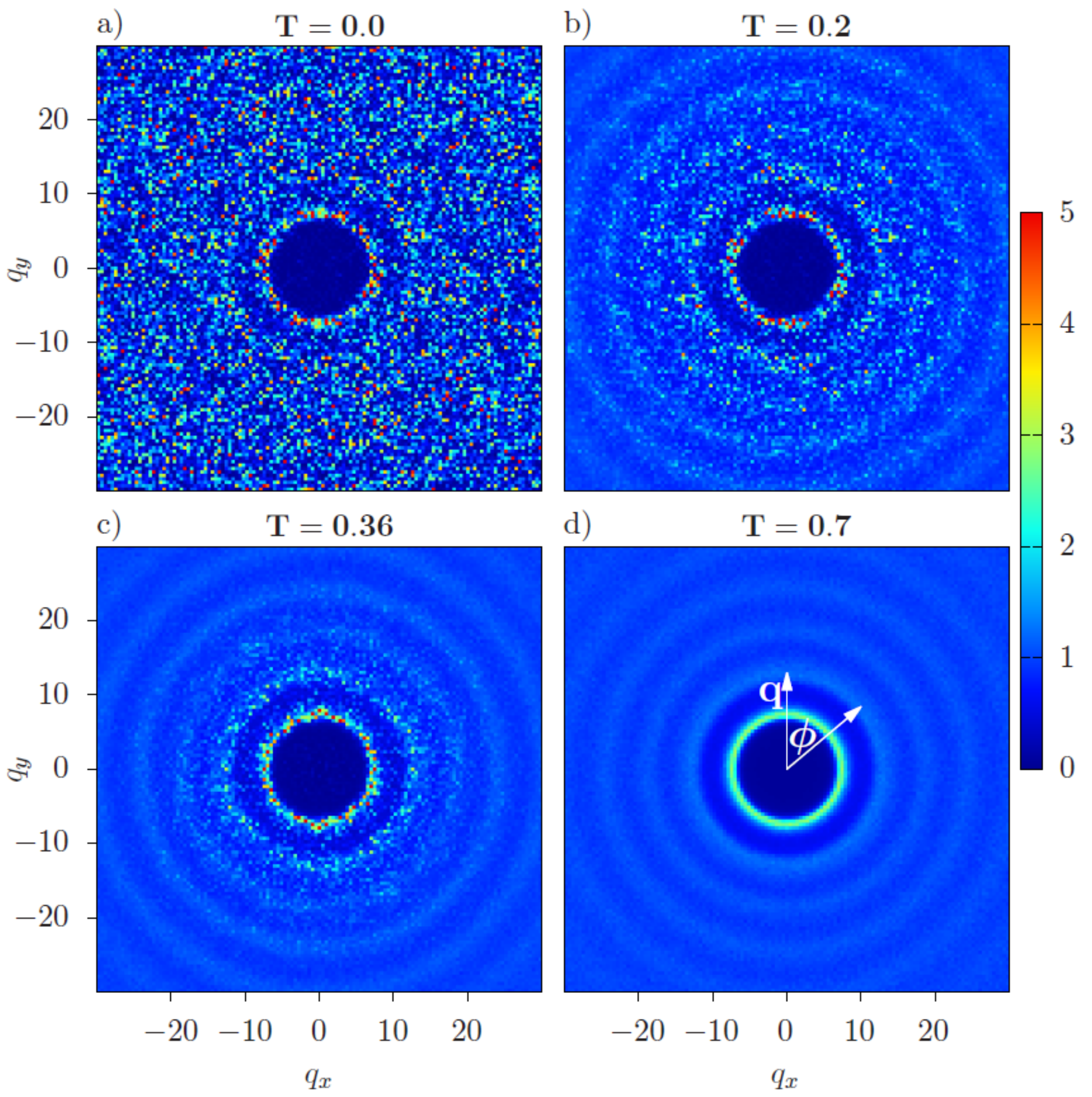

**Figure 1.** The calculated scattering pattern for a range of temperatures using data from a simulated binary Lennard-Jones mixture described in Section 2 averaged over a measurement time $\tau = 2 \times 10^4$. The disappearance of the angular variance of the scattering intensity on heating is clearly evident. There is weak scattering within the radius of the inner ring. The apparent absence of scattering at small q is a result of the color scale being unable to resolve the low scattering intensity in this region.

These existing methods address the question of structure either in terms of the length scale of medium range order [15] or in the spatial variation of relaxation kinetics [16]. As described above, we are interested in a different approach to structure, one based, not on characterising a structure by the positions of particles, but by the capacity of a configuration to restrain the motion of the constituent particles. In 2023, we reported [17] a method for measuring restraint using a simple linear relationship between the scattering intensity at a temperature T and at zero temperature of the form

$$< S(\vec{q},T)-1 >_{S_o} = D_w(q,T)[S(\vec{q},0)-1]+H(q,T) \tag{3}$$

where $<\ldots>_{S_o}$ indicates a restricted average of that set of wavevectors associated with a given value of $S(\vec{q},0)-1$. Eq. 3 was derived assuming Gaussian distributions of particle displacements. The quantity $D_w(q,T)$ was shown to be in reasonable agreement with the Debye-Waller factor $D_o(q,T)$,

$$D_o(q,T) = \left\langle exp\left(-\frac{q^2}{3}\left\langle\left(\vec{r}_j - \overline{r}_j\right)^2\right\rangle_\tau\right)\right\rangle_j \tag{4}$$

where the average $\langle \dots \rangle_\tau$ is an average over the same time interval τ used to measure the intensities (see Eq. 1), $\overline{r_j} = \langle \vec{r_j} \rangle_\tau$ and $\langle \dots \rangle_j$ is the average over the particles in the system. In addition, the quantity H(q,T) provided a measure of the spatial variation in the restraint experienced by individual atoms [17]. While successful when tested on a simulated glass-former, our 2023 proposal for the analysis of the temperature dependence of scattering is difficult to implement experimentally. The method requires that the intensity at each wavevector be tracked over an extended observation time during which a range of temperatures are sampled. Such measurements would be extremely sensitive to shifts in the orientation of the sample in the beam, decoherence of the radiation source and radiation-induced degradation of the material [18,19]. To address this problem, we present a new methodology for the analysis of coherent scattering intensities that avoids the need for extended monitoring of individual reflection intensities. Since the scattering at each value of the control parameter (i.e. temperature, density, etc) is treated independently, the methodology presented below can be equally applied to non-thermal control parameters such as concentration in the case of colloidal glass formers.

The paper is divided into two parts. In Section 2, we define the restraint R(q,T) and calculate its variation with respect to temperature. We then compare R(q,T) with two properties that are expected to be strongly correlated with any reasonable measure of atomic restraint, namely the Debye-Waller factor and the shear modulus. In Section 3 we consider some practical issues regarding the experimental implementation and interpretation of the restraint. Here we consider the dependence of R(q,T) on the choice of wavevector q and how sensitive the restraint is to the degree of coarse graining over the reciprocal space used when measuring scattering intensity.

**2. Structural Restraint and its Correlation with the Debye-Waller Factor and the Shear Modulus**

The structure factor, as shown in Fig.1, is presented as a distribution of intensity in the $(q_x,q_y)$ plane obtained by setting $q_z = 0$. We can change the coordinates of these plots from the Cartesian $(q_x,q_y)$ to the circular $(q, \phi)$, as depicted in Fig.1d. Averaging the intensities over the angle $\phi$ gives us the standard liquid structure factor $\overline{S}(q,T)$,

$$\overline{S}(q,T) = \langle S(\vec{q},T) \rangle_{\phi} \ , \tag{5}$$

At this point, it is useful to consider the status of $\overline{S}(q,T)$. As defined in Eq.5, the standard liquid structure factor can be obtained by explicitly carrying out the averaging of intensities over the angle $\phi$. In practice, the same effective averaging occurs when measuring the scattering using an incoherent radiation source, i.e. without the need for explicit angle averaging. The speckle pattern, on which our measure of structural restraint depends, requires a coherent radiation source such as x-rays from a synchrotron source, electron beams and, in the case of scattering from colloidal suspensions, standard optical lasers.

The model system we are using to generate the atomic configurations is the binary $A_{80}B_{20}$ model glass former introduced by Kob and Andersen (KA) [20]. The molecular dynamics simulations are performed at fixed NVT with the LAMMPS software [21], using a Nose-Hoover thermostat to keep the temperature constant. All quantities are reported in reduced Lennard-Jones units as described in Ref [20]. Our system consists of a binary mixture of 5000 particles at a density of 1.20. Error bars are calculated from 8 independent repetitions of the simulations. Further details of the calculations can be found in ref.[17].

We have chosen a measurement time $\tau = 2 \times 10^4$. The variance in the speckle pattern depends on the choice of $\tau$ since, at any temperature, the pattern can be lost by averaging over a sufficiently long time. In practice, however, the observation time for the measurement of scattering intensity is bounded below by the minimum time required to sample sufficient photon count (a function of beam intensity) and is bounded above by the stability of the sample and the beam. Within this range the variation in the characteristic temperature for the onset of restraint is small, given the rapid increase in the relaxation time with decreasing temperature. In molecular dynamics simulations we are restricted to very short observation times and so will see the onset of restraint at a higher temperature. While this limitation restricts quantitative comparison between experiment and simulation, it does not affect our goal in this paper of demonstrating the viability of measuring structural restraint using the variance of the scattered intensity.

In Fig. 2 we plot the structure factor $\overline{S}(q,T)$ as a function of q, the magnitude of the wavevector, for a range of temperatures. The first peak in the structure factor occurs at $q_1 = 7.2$. To get some sense of the range temperatures that we cover in this study, we can identify three characteristic (reduced) temperatures for the KA model: the freezing point $T_f = 1.03$ [22], the mode coupling temperature $T_{MC} = 0.435$ [23] and the glass transition tempertaure $T_g \sim 0.32$ [23].

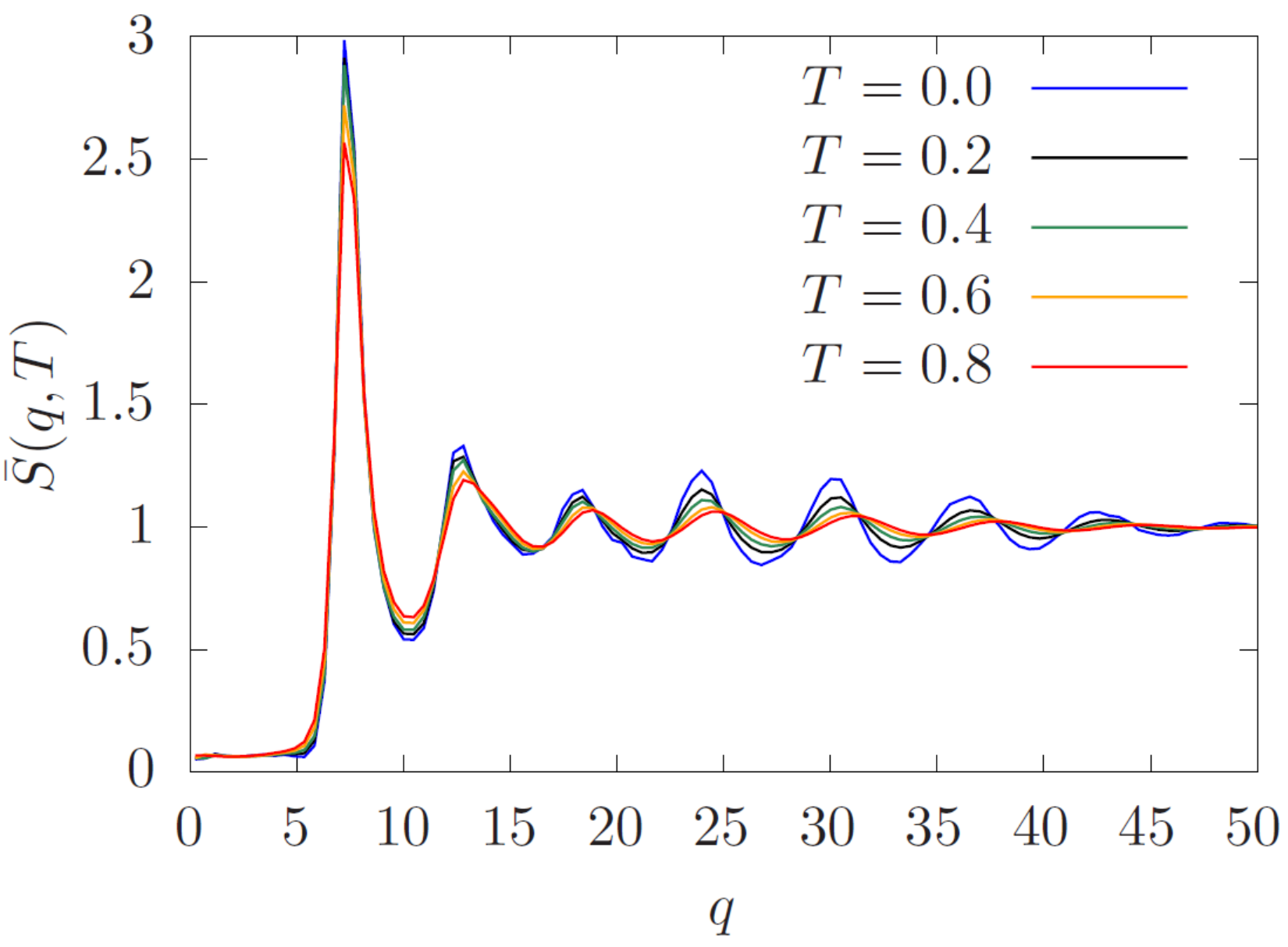

**Figure 2.** The structure factor $\bar{S}(q,T)$ as a function of q, the magnitude of the wavevector, for a number of temperatures, as indicated. The first peak in $\bar{S}(q,T)$ occurs at $q_1 = 7.2$..

We now define our order parameter of structural restraint R(q,T) as

$$R(q,T) = \frac{\left\langle \left( S(\vec{q},T) - \bar{S}(q,T) \right)^2 \right\rangle_{\varphi}}{\left\langle \left( S(\vec{q},0) - \bar{S}(q,0) \right)^2 \right\rangle_{\varphi}} \tag{6}$$

We choose $q = q_1$ because it corresponds to the largest average scattering intensities. We will explore other choices in Section 3. The use of the T = 0 scattering as a reference is a convenient choice in our simulations but is not essential. What is required is that the reference variance is measured at a temperature low enough that the sample is regarded as a solid. In other words, the choice of reference essentially represents an operational definition of solidity. As discussed above, the restraint measure R(q,T) does not require the extended tracking of the intensities of individual reflections (as had been proposed in ref. 17) but only the variance of the intensity distribution over the angle $\phi$ at a given value of q. We plot $R(q_1,T)$ vs T in Fig.3. The comparison with the ratio $\frac{\bar{S}(q_1,T)}{\bar{S}(q_1,0)}$ of the peaks of the structure factors at T and T = 0 are also plotted for comparison. The reduced intensity $\frac{\bar{S}(q_1,T)}{\bar{S}(q_1,0)}$ shows only a small continuous variation over the temperature range. Mauro et al [24] have reported the observation of an 'acceleration' in the increase of $\bar{S}(q,T)$ in X-ray scattering from fragile glass-forming metal alloys during cooling from just above the glass transition temperature. They linked this feature with structural ordering associated with the fragile kinetics. As shown in Fig. 3, we do not find this feature in the binary Lennard-Jones mixture.

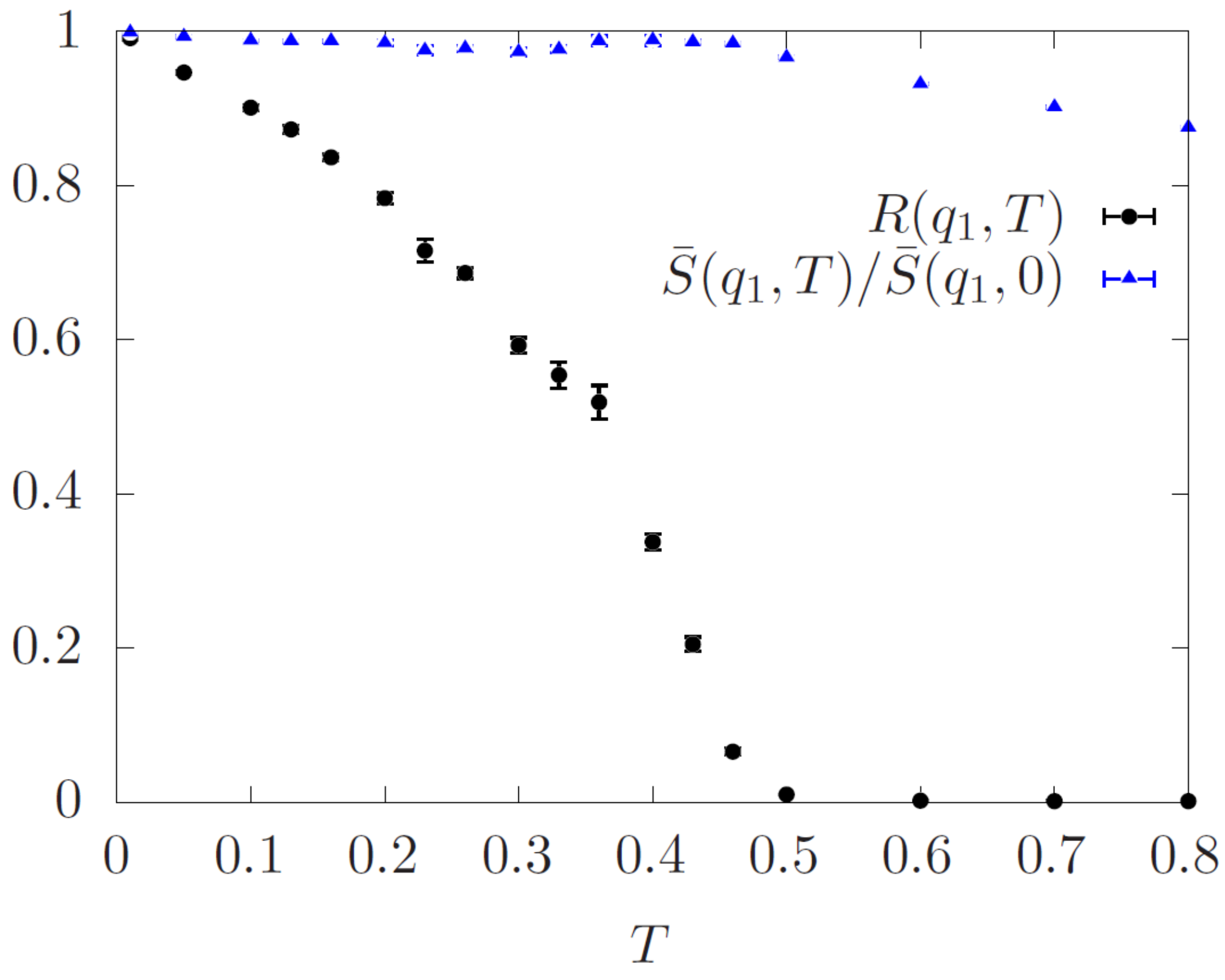

**Figure 3.** Plot of the restraint order parameter R(q$_1$,T) vs the temperature T. For comparison, we also plot $\frac{\bar{S}(q_1,T)}{\bar{S}(q_1,0)}$ . Note the data at q$_1$ consists of an average over q$_1$-0.1 < q < q$_1$+0.1.

The utility of an order parameter rests primarily on its capacity to differentiate structurally the low and high temperature states and to do so in such a way as to correlate with our perception of the distinguishing physical properties. So, how does the structural restraint R(q$_1$,T) fare? As shown in Fig. 3, the restraint R(q$_1$,T) exhibits a substantial variation between low and high temperatures, satisfying our first requirement. With regards to the second, we shall consider two properties associated with the generic differentiation of solid and fluid states. The first is the Debye-Waller factor $D_o\left(q_1,T\right)$, already defined in Eq.4, a quantity closely associated with atomic restraint. The second property we consider is the shear modulus G(T) of the sample calculated using the fluctuation formula of Squire, Holt and Hoover [25],

$$G(T) = G_\infty(T) - \beta V \left\langle \left(\sigma - <\sigma>_\tau\right)^2 \right\rangle_\tau$$

where $\sigma$ is the instantaneous shear stress of the system which is defined [26] as

$$\begin{aligned} \sigma &= \frac{1}{V}\frac{dU}{d\gamma}\Big|_{\gamma=0} = \frac{1}{V}\sum_{i<j}\sigma_{\alpha\beta}(\vec{r}_{ij}) \\ \sigma_{\alpha\beta}(\vec{r}_{ij}) &= \frac{d\phi_{\alpha\beta}(r_{ij})}{d\gamma}\Big|_{\gamma=0} = r_{ij}\phi'_{\alpha\beta}(r_{ij})\hat{x}_{ij}\hat{y}_{ij} \end{aligned} \tag{7}$$

where the shear strain $\gamma$ is assumed to be in the x-z plane, i.e. $\Delta x_i = \gamma z_i$ , $\hat{x}_{ij} = (x_i - x_j)/r_{ij}$ and $\phi_{\alpha\beta}(r)$ is the interaction potential between species α and β. The infinite frequency modulus G$_\infty$ is given by

$$G_\infty = \frac{N}{V}k_B T - \frac{1}{2V}\sum_i\sum_{j\neq i}\left\langle \left( y_{ij}^2 F_{ij}\left[1 - \frac{x_{ij}^2}{r_{ij}^2}\right] - \frac{d^2\phi_{\alpha\beta}}{dr_{ij}^2}\frac{x_{ij}^2 y_{ij}^2}{r_{ij}^2}\right)\right\rangle \tag{8}$$

where $F_{ij} = -\left(1/r_{ij}\right)d\phi_{\alpha\beta}/dr_{ij}$ and the average in Eq. 8 is over the time interval τ

The variance and mean of the shear stress is, in the results presented here, calculated over the same measurement time τ as the restraint and the Debye-Waller factor. This inclusion of an implicit time dependence in the calculation of the shear modulus has been examined previously [27]. In Fig. 4, we plot the calculated values of the Debye-Waller factor and the shear modulus, the latter normalised by its T=0.01 value.

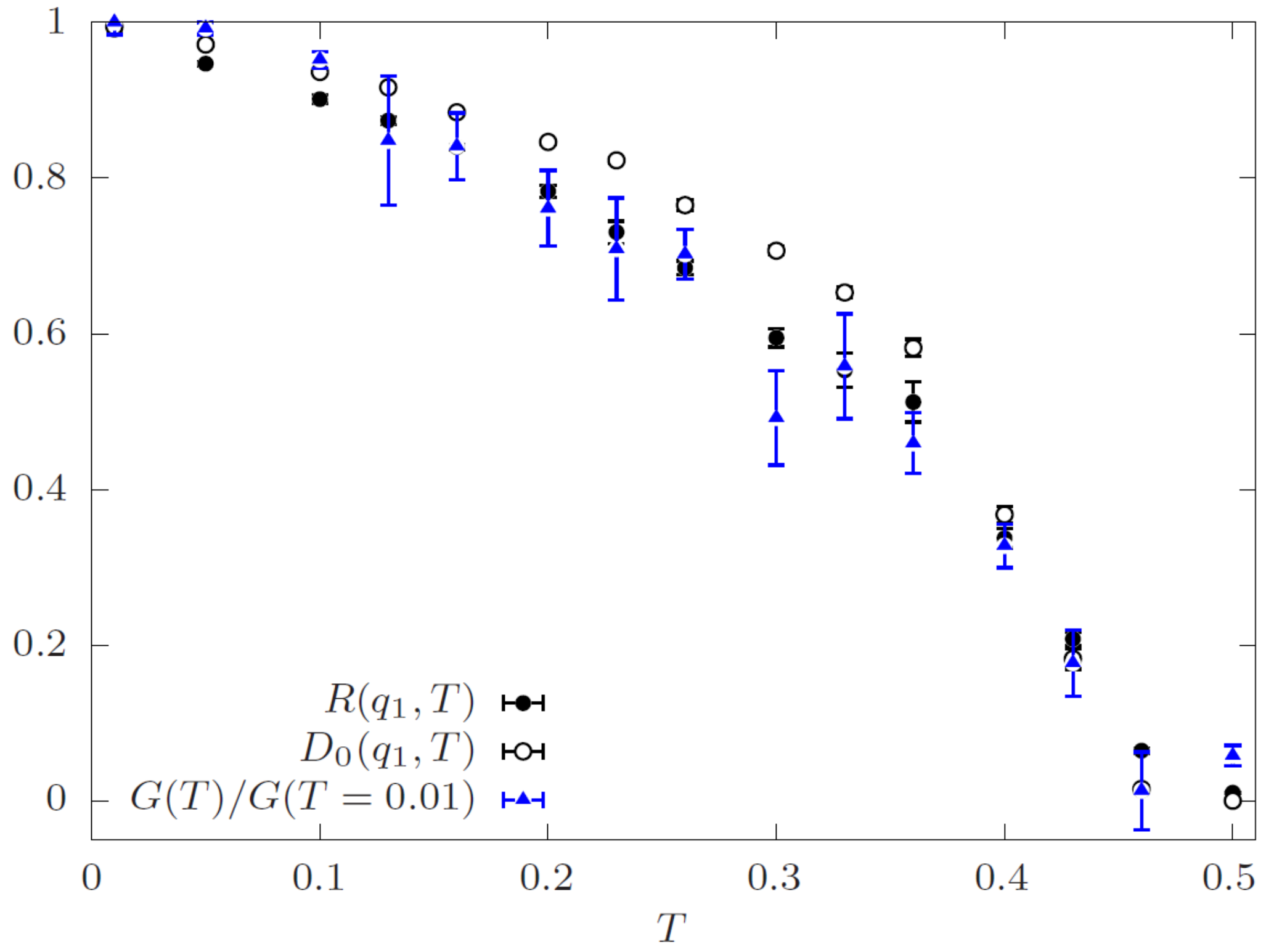

**Figure 4.** The plot of $R(q_1,T)$, $D_o(q_1,T)$ and the reduced shear modulus $G(T)/G(T=0.01)$ as a function of T. All quantities are calculated using the same choice of measurement time.

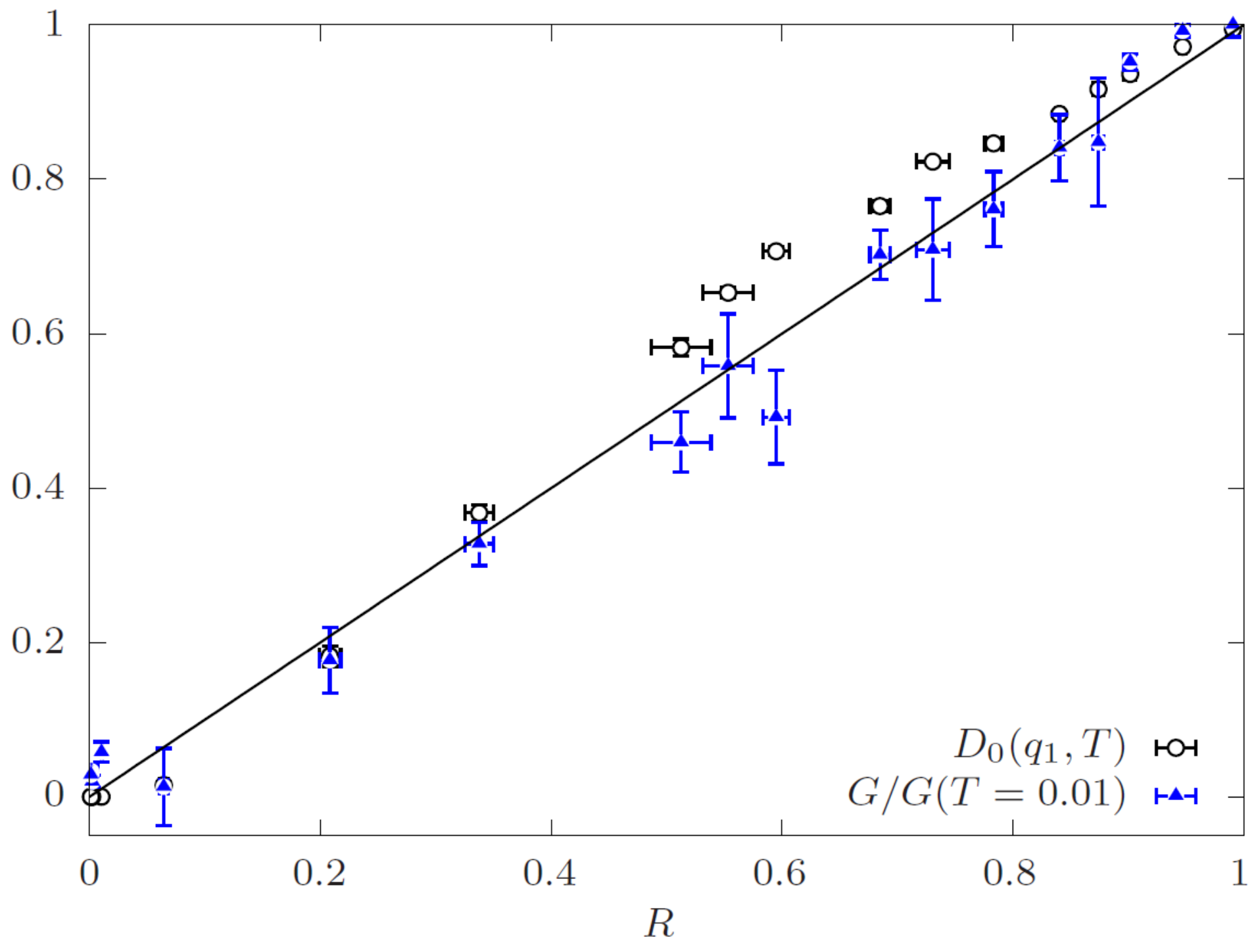

**Figure 5.** A plot of the Debye-Waller factor $D_o(q_1,T)$ and the reduced shear modulus G(T)/G(T=0.01) vs the restraint $R(q_1,T)$.

As shown in Fig. 4, the variation of $R(q_1,T)$ closely matches that of the reduced shear modulus G(T)/G(0.01). The agreement between $R(q_1,T)$ and the Debye-Waller (DW) factor $D_o(q_1,T)$ is not quite so good. We find a systematic positive deviation of the DW factor from the restraint over the important temperature range $0.15 < T < 0.35$. These comparisons are perhaps more easily represented when the reduced shear modulus and the DW factor are plotted directly against the restraint in Fig. 5. Our motivation for presenting this comparison was to test the proposition that the restraint represents a reasonable order parameter for quantifying the structural transition between fluid and amorphous solid. The results plotted in Figs. 4 and 5 certainly support that proposition. While the observation that both quantities share a monotonic relation with $R(q_1,T)$ is to be expected, the agreement we find between the temperature dependence of the restraint and the shear modulus is surprising, especially as the latter quantity involves highly nonlinear coupling between the fluctuations in atomic positions and shear stress. We cannot, at this stage, account for this agreement nor make any prediction about its generality but the possibility of being able to determine how a shear modulus varies with temperature from a scattering measurement is intriguing.

### 3. On the Influence of Wavevector and Angular Coarse Graining

Our goal in this paper is to propose a practical method for the experimental measurement of atomic restraint in condensed matter. As already discussed, $R\left(q_1,T\right)$, our measure of structural restraint, has been specifically designed to avoid the practical difficulties of monitoring the temperature dependence of the intensity of individual points in reciprocal space. In this Section, we shall consider how two implementation choices influence the measured values of the structural restraint. These choices are i) the magnitude of the scattering wavevector q, and ii) the coarse graining of the intensities.

The virtue of choosing $q_1$, the value corresponding to the first peak in the structure factor, is that this choice makes use of the most intense scattering and so improves resolution of variation in intensity associated with the speckle. As this wavevector might not be accessible in a given experiment, it is useful to understand how the measured restraint varies with q. To this end, the values of R(q,T) vs T for a range of q values have been plotted in Fig. 6. We find that the choice of q shifts the temperature range over which R transitions between high and low values. To quantify this dependence, we can define a characteristic temperature $T_c$ as the value at which R(q,T) = 0.5 and plot the dependence of $T_c$ on q (see Fig. 7). As shown in the Fig. 7 inset, the variance in the scattering intensity drops dramatically when q decreases below a value of ~ 5, making the calculation of R(q,T) for smaller wavevectors unreliable. On inspection of Fig. 7, we find that the value of $T_c$ exhibits a sharp maximum at $q = q_1$, indicating that the onset of structural restraint on cooling is associated with density fluctuations on a length scale set by the Bragg wavevector.

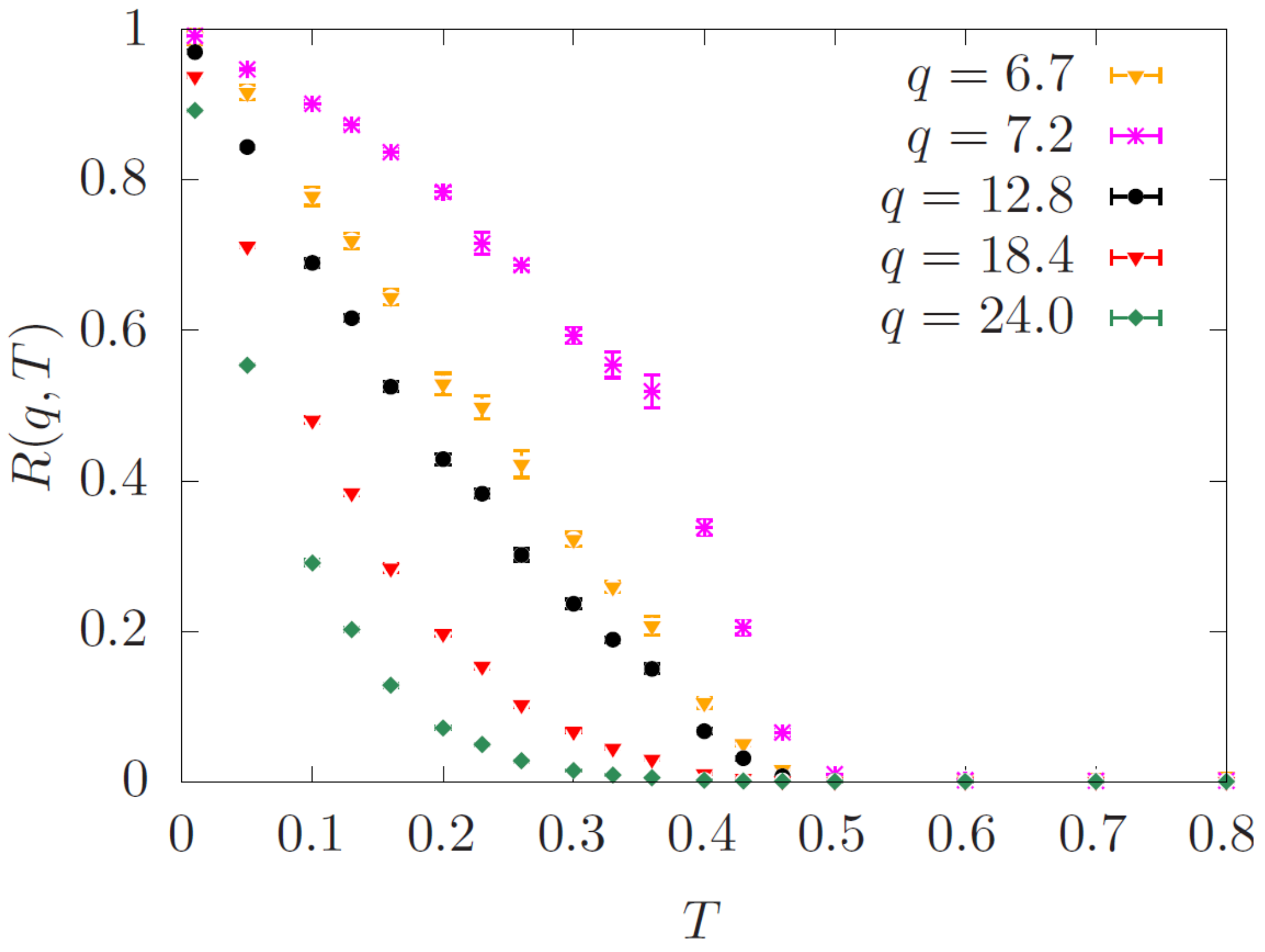

**Figure 6.** Plot of the restraint order parameter $R(q_1,T)$ vs the temperature T for a range of values of the wavevector q. Note the data at a given q consists of an average over q - 0.1 < q < q + 0.1.

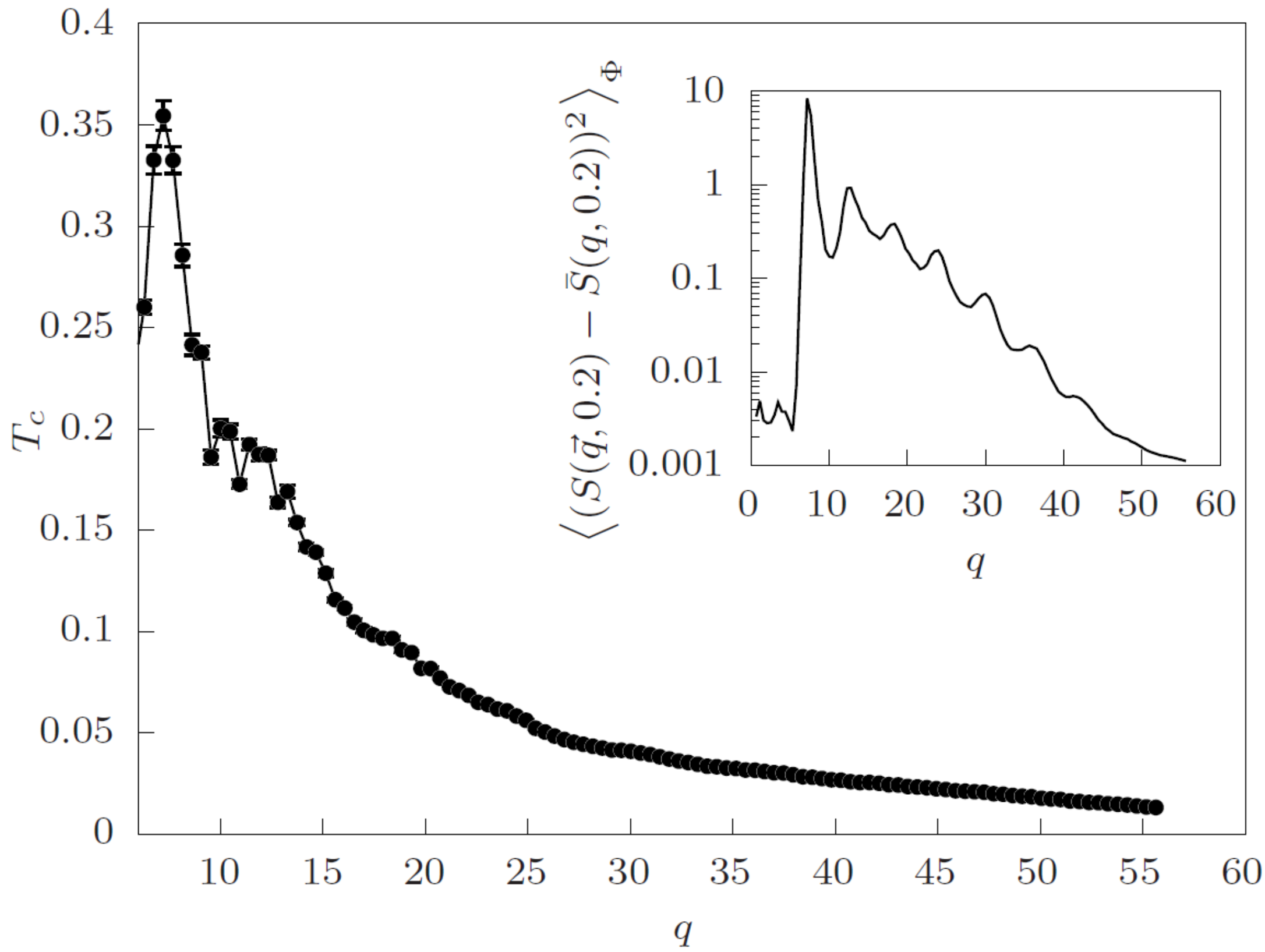

**Figure 7**. The dependence of the characteristic temperature $T_c$ (defined in text) for the onset of structural restraint on the choice of the scattering vector q. Inset. The variance in the scattering intensity as a function of q. Note the abrupt drop in the variance for q < 5.

Finally, we consider the influence of coarse graining the collection of intensity on the measurement of R. The scattered intensity is typically collected as photon counts per pixel on a 2D detector. Depending on the beam intensity and distance of the detector from the sample, the count per pixel might be small, possibly just 1 or zero. In such cases, any variation in intensity due to the structure of the sample is completely washed out by the fluctuations due to shot noise. To extract useful information from such measurements, we need to combine intensity over a group of pixels. This unavoidable coarse graining will reduce the variance in the intensity fluctuations, leaving us with the question, how sensitive is $R\left(q_1,T\right)$ to the degree of coarse graining used in collecting intensities?

To mimic the kind of pixel averaging required experimentally, we have defined an angular coarse-grained pixel in terms of a segment defined by an angle δ and, radially, an annulus of width d. In Figs. 8 and 9 we have plotted the variance $\left\langle\left(S\left(\vec{q}_1,T\right)-\bar{S}\left(q_1,T\right)\right)^2\right\rangle_\phi$ and structural restraint R(q,T) as a function of a coarse graining angle δ for T = 0.2 and 0.36, respectively. As expected, the variance of the speckle intensity exhibits a steep monotonic decrease as we increase the angular extent of the coarse-grained pixel. The structural restraint, a ratio of two variances, shows no such systematic dependence on δ. As the variance $\left\langle\left(S\left(\vec{q}_1,T\right)-\bar{S}\left(q_1,T\right)\right)^2\right\rangle_\phi$ decreases with the increase in the coarse-grained pixel size, the structural restraint exhibits increasing statistical noise, the result of taking the ratio of two small numbers. However, for coarse graining that leaves the structure factor variance > 0.1, we find that $R\left(q_1,T\right)$ exhibits little variation with coarse graining. We conclude that it should be possible to obtain a consistent value of the structural restraint over a range of coarse graining that retains sufficient speckle structure.

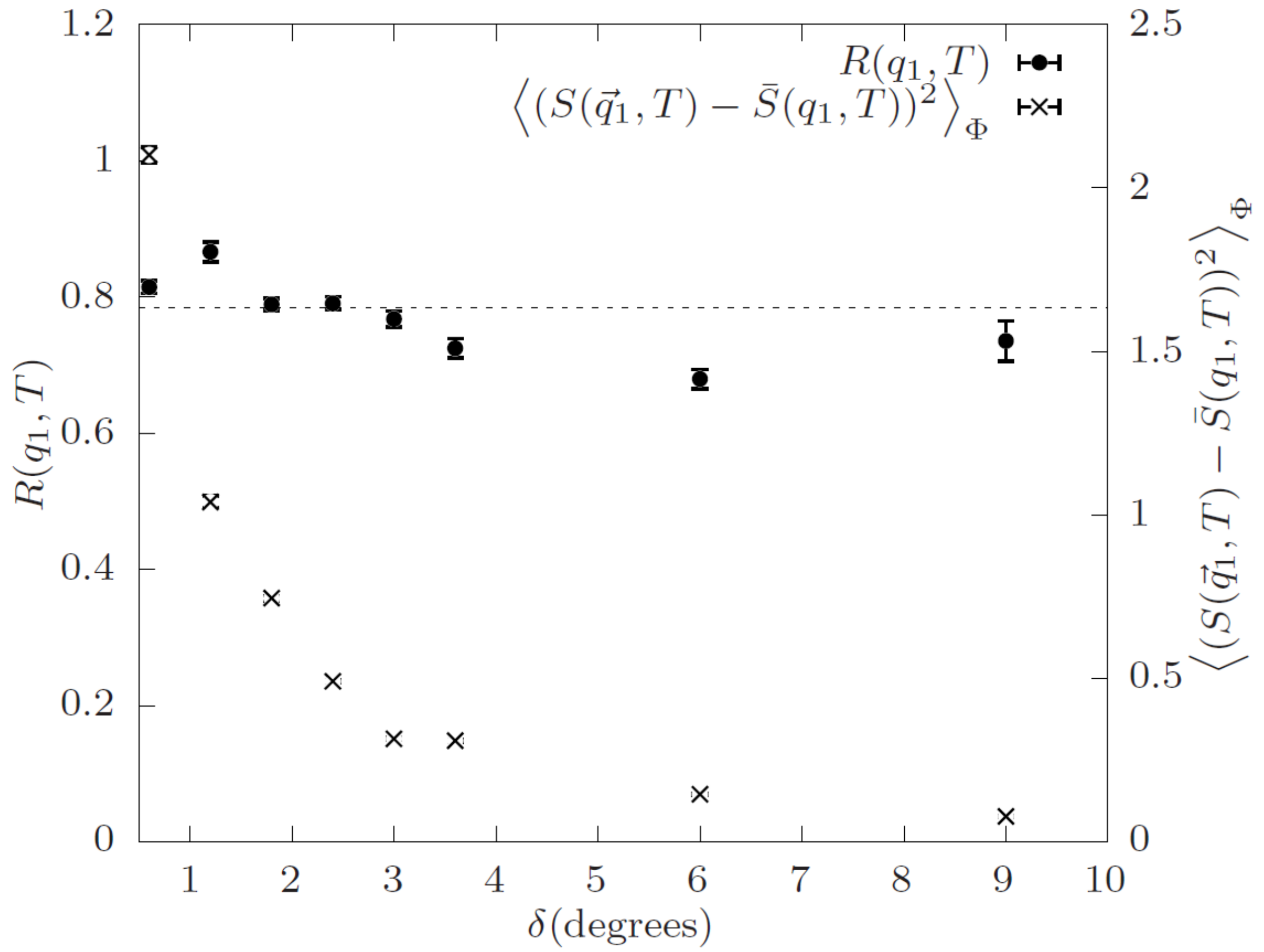

**Figure 8.** A plot of the structural restraint $R(q_1, T)$ and the variance $\left\langle \left(S(\vec{q}_1, T) - \bar{S}(q_1, T)\right)^2 \right\rangle_\phi$ as a function of the angular size δ of the coarse graining volume for T = 0.2.

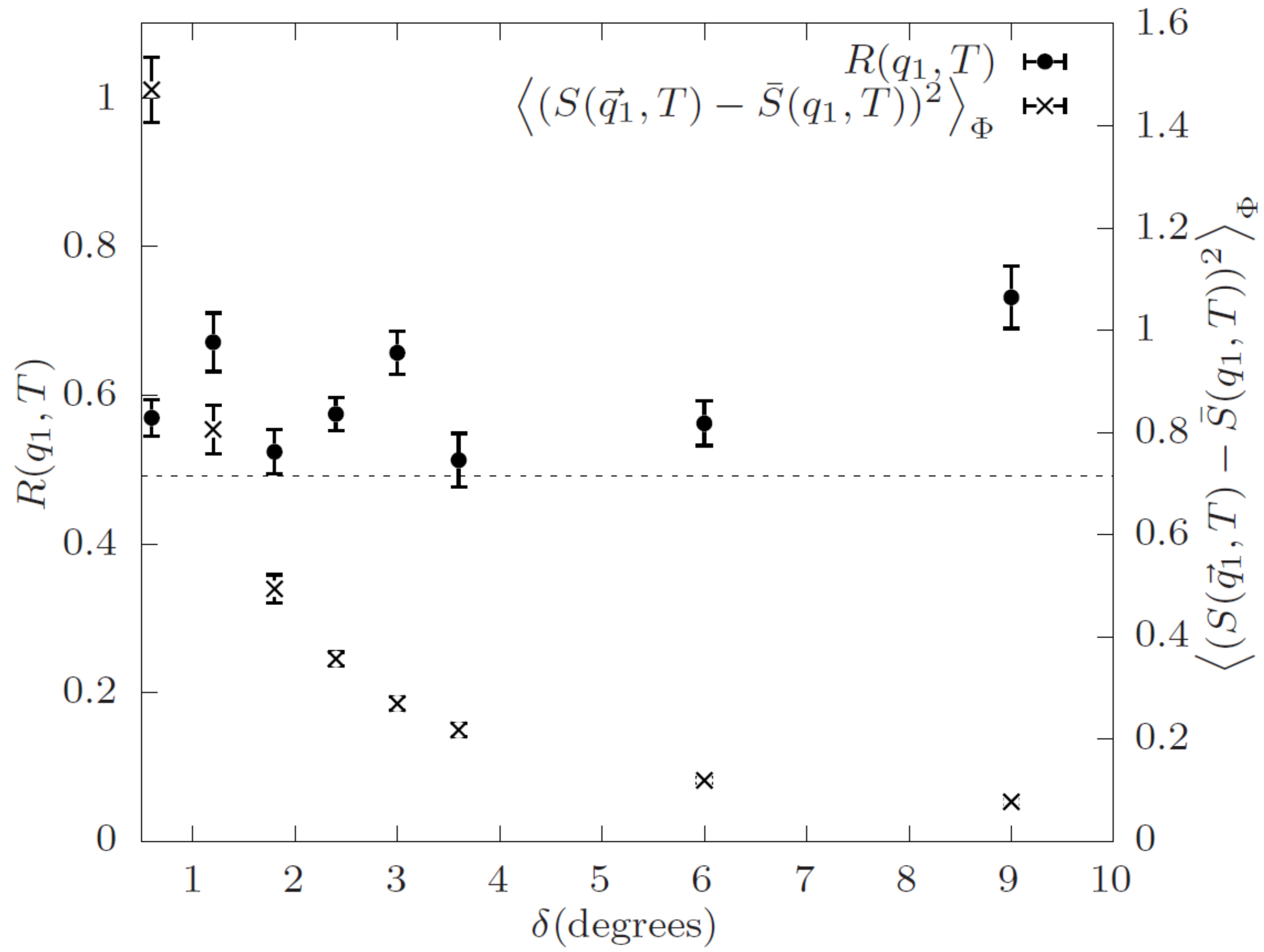

**Figure 9.** A plot of the structural restraint $R(q_1,T)$ and the variance $\left\langle \left(S(\vec{q}_1,T) - \bar{S}(q_1,T)\right)^2 \right\rangle_\phi$ as a function of the angular size δ of the coarse graining volume for T = 0.36.

## 4. Conclusions

We have presented an experimentally accessible measure of structural restraint, the quantity R(q,T), defined as the ratio of the angular variance of the scattering intensity over the analogous variance at a temperature sufficiently low to provide an operational solid. Choosing the wavevector to equal that of the first peak of the liquid structure factor, $q_1$, the structural restraint identifies the onset temperature for the appearance of solid-like atomic restraint from elastic scattering alone. While we have used temperature as our control parameter in this study, it is straightforward to extend our definition of the order parameter to situations where the control parameter is density (as in the case of colloidal glasses) or applied strain (as in mechanical devitrification).

The restraint R(q,T) represents a general and, we believe, useful tool for experimentally characterising that aspect of structural change that most directly influences the mechanical properties of disordered materials. This contention is supported by the striking alignment of $R\left(q_1,T\right)$ and the reduced shear modulus reported here. We note that the method is not restricted to states at equilibrium, stable or metastable. Scattering from amorphous samples that have been subjected to mechanical strain [28] or irradiation-induced damage and materials undergoing chemical transformations can be characterised by this approach.

**Acknowledgements**

We gratefully acknowledge valuable conversations with Beatrice Ruta, Giacomo Baldi and Amelia Liu. C.F.P. is the recipient of an Australian Research Council Discovery Early Career Award (No. DE210100256) funded by the Australian Government. This research was undertaken with the assistance of resources and services from the National Computational Infrastructure (NCI), which is supported by the Australian Government.